\documentclass[twocolumn,amsmath,amssymb,aps,]{revtex4-1}
\usepackage{bm}
\usepackage[colorlinks=true,urlcolor=blue,linkcolor=blue,citecolor=blue]{hyperref}
\usepackage{color}
\usepackage{graphics}
\usepackage{graphicx}
\usepackage{epsfig}
\usepackage{amssymb}
\usepackage{amsmath}
\usepackage{hyperref}
\usepackage{physics}

\usepackage{array}                 

\usepackage{amsfonts}           

\begin{document}

\title{Penning-trap eigenfrequency measurements with optical radiofrequency detectors}

\author{J.~Berrocal$^{1}$} \email[]{This is part of the PhD thesis of Joaqu\'in Berrocal.}
\author{A.~Hern\'andez$^1$}
\author{I.~Arrazola$^2$}
\author{F.~Dom\'inguez$^1$}
\author{A.~Carrasco-Sanz$^{3}$}
\author{F.J.~Fern\'andez$^4$}
\author{M.~Block$^{5,6,7}$}
\author{D.~Rodr\'iguez$^{1,8}$}\email[]{danielrodriguez@ugr.es}
\affiliation{
$^1$Departamento de F\'isica At\'omica, Molecular y Nuclear, Universidad de Granada, 18071 Granada, Spain \\
$^2$Vienna Center for Quantum Science and Technology, Atominstitut, TU Wien, 1040 Vienna, Austria\\
$^3$Departamento de \'Optica, Universidad de Granada, 18071 Granada, Spain\\
$^4$Departamento de Arquitectura y Tecnolog\'ia de Computadores, Universidad de Granada, 18071 Granada, Spain\\
$^5$Department Chemie - Standort TRIGA, Johannes Gutenberg-Universit\"at Mainz, D-55099, Mainz, Germany \\
$^6$GSI Helmholtzzentrum f\"ur Schwerionenforschung GmbH, D-64291, Darmstadt, Germany \\
$^7$Helmholtz-Institut Mainz, D-55099, Mainz, Germany\\ 
$^8$Centro de Investigaci\'on en Tecnolog\'ias de la Informaci\'on y las Comunicaciones, Universidad de Granada, 18071 Granada, Spain}

\date{\today}

\begin{abstract}
We use an electric-dipole laser-driven transition to precisely measure the cyclotron-frequency ratios of the pairs $^{42}$Ca$^+$-$^{40}$Ca$^+$, $^{44}$Ca$^+$-$^{40}$Ca$^+$ and $^{48}$Ca$^+$-$^{40}$Ca$^+$ in a 7-tesla Penning trap. A single laser-cooled ($T\approx 1$~mK) ion serves, together with photon-counting and/or photon-imaging units, as a radiofrequency detector covering a broad-band frequency spectrum, in the present case from kHz to a few MHz. Such detectors ($^{40,42,44,48}$Ca$^{\scriptsize{+}}$) allow measuring extremely small forces, with measured normalized sensitivities down to $7.4(3.5)$~yN$/\sqrt{\text{Hz}}$ and $24.9(9.9)$~yN$/\sqrt{\text{Hz}}$ in the MHz and kHz regime, respectively. The direct determination of the ions' amplitudes makes a cyclotron-frequency measurement process more robust against inhomogeneities of the magnetic field and/or deviations of the electric quadrupole field due to mechanical imperfections of the trap.
\end{abstract}


\maketitle

Penning traps have a distinct and unique feature such as the natural presence of a large magnetic field, enabling the stable detection of extremely weak forces and electric fields with many trapped ions \cite{Bier2010,Gilm2021}. In addition, the accurate determination of the cyclotron frequency of a charged (anti)particle confined in the trap can be utilized to evaluate fundamental constants \cite{Myer2019}, to perform fundamental symmetries tests \cite{Ulme2015,Myer2018} and to search for new Physics \cite{Rain2004,Rain2005,Myer2015,Schu2020}. Penning-trap single-particle detectors have been developed on the basis of electronic detection following the principle of the stored ion calorimeter \cite{Wine1975}. The cyclotron frequency
\begin{equation}
\nu _c =\frac{1}{2\pi}\frac{q}{m}B, \label{eq:cyclotron}
\end{equation}
directly proportional to the particles' mass-to-charge ratio ($q/m$), is determined from the measurements of three characteristic frequencies of the target particle \cite{Brow1986}. A measurement of the magnetic field strength ($B$) is avoided by measuring cyclotron-frequency ratios, i.e., alternating the $\nu _c$-measurement of the target ion with the $\nu _c$-measurement of another charged particle, with well-known $q/m$ \cite{Rain2004,Fili2021}. In single-ion Penning-trap experiments carried out to date, the three characteristic frequencies of the ion are determined by damping one of the ion's eigenmotions through a resonant tank circuit at 4~K  \cite{Weis1988,Ulme2009}. Thus, one eigenfrequency is obtained directly by Fourier transformation of the induced-current signal, and the remaining ones are obtained in the same way after applying motional conversion \cite{Rain2004,Corn1990}. 
In this paper we report on direct measurements of the three characteristic frequencies of a single laser-cooled calcium ion in a 7-tesla Penning trap, utilizing the ion as radiofrequency (rf) fluorescence detector \cite{Wine1983} covering the range from kHz to MHz. We present the cyclotron-frequency ratios of $^{42}$Ca$^+$-$^{40}$Ca$^+$, $^{44}$Ca$^+$-$^{40}$Ca$^+$ and $^{48}$Ca$^+$-$^{40}$Ca$^+$ from the observation of the fluorescence photons of the $^2$S$_{1/2}\rightarrow ^2$P$_{1/2}$ electric-dipole transition in calcium after applying external rf fields. Precise  cyclotron-frequency ratios of even calcium isotopes have been used, combined with the mass value of  $^{40}$Ca \cite{Nagy2006} and spectroscopy data \cite{Gebe2015,Knol2019,Sola2020}, to probe fundamental symmetries, and there is still room for improvement.

For a Doppler cooled ion, a decrease in its temperature from 4~K (electronic detection) to 1~mK (optical detection), directly translates into a reduction of the ion's oscillation amplitude. 
It is thus sensitive to extremely small forces, and the measurement will be less affected by magnetic field inhomogeneities, imperfections of the electric field, relativistic effects, and induced charges on the trap electrodes. Furthermore, the ion is (optically) observed before and after the external field is applied.
This ion-based detector can be utilized in accurate mass spectrometry on any atomic or molecular ion, which cannot be directly laser cooled, when the target and calcium ion are confined in the same potential well of the Penning trap, forming an unbalanced two-ion Coulomb crystal \cite{Guti2019b,Cerr2021}. Such a configuration can be used for identification of exotic single particles, such as ions from superheavy elements, complementing the measurements carried out to date and permitting the extension of the survey in this mass region \cite{Bloc2010,Mina2012,Kale2022}. 
\begin{figure*} [t!]
\includegraphics[scale=0.50]{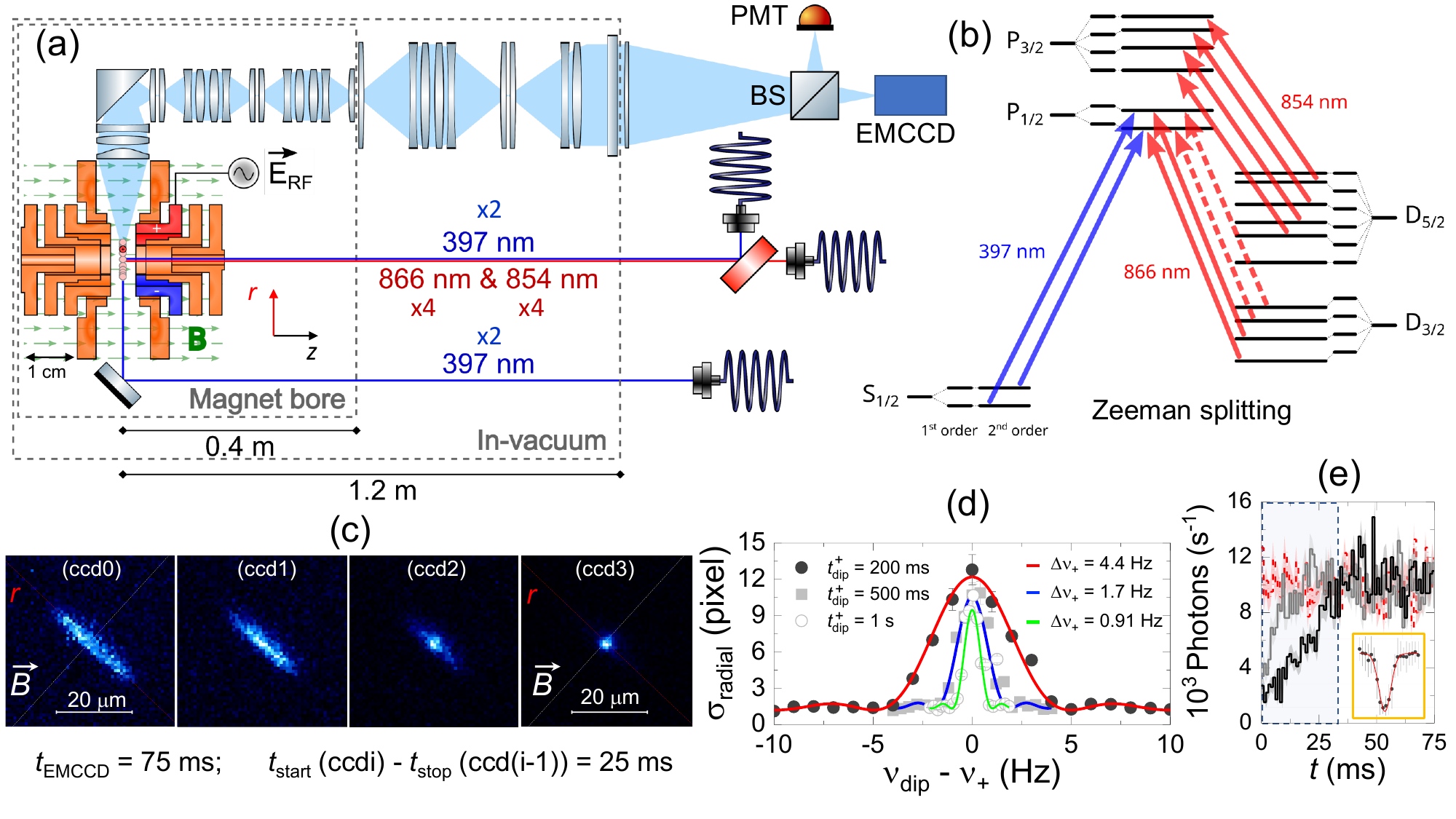}
\vspace{-7mm}
\caption{\label{figure1} a) Sketch showing a longitudinal cut of the open-ring Penning trap \cite{Guti2019,Berr2022}, the dipolar electric field (in the radial direction), and the optical system (not to scale) for photon detection. b) Level scheme of calcium ions in a strong magnetic field including first and second-order Zeeman splittings, indicating all the transitions to be driven to perform Doppler cooling. c) Sequence of EMCCD images of  a calcium ion on applying Doppler cooling after probing the modified-cyclotron motion with an external field oscillating at $\nu _{\text{dip}}=\nu _+$.  d) Standard deviation of the Lorentzian fit to the projections along the $r$ direction of EMCCD images [e.g., (c)-ccd0], versus frequency of the external field for three different excitation times.  $\Delta \nu$ corresponds to the FWHM of the fitted Rabi sinc function. e) Number of photons registered by the PMT for three different values of $\nu _{\text{dip}}$; when the oscillating electric field is not resonant with the axial motion (dotted red), when $\nu _{\text{dip}}=\nu _z$ (solid black) and when $\nu _{\text{dip}}=\nu _z+\Delta \nu _z/2$ (solid grey). The inset shows the number of photons collected within the time window delimited by the dashed lines versus frequency of the external field. }
\end{figure*}
Motional frequency metrology on any ion species will be also practicable provided the Coulomb crystal is cooled to the ground state of motion \cite{Cerr2021}.

In the Penning trap, the motion of an ion with a given $q/m$ in the homogeneous magnetic field ($\sim 0.1$~ppm within 1~cm$^3$), and under an ideally perfect quadrupolar electrostatic field is depicted as the superposition of three eigenmotions, one in the axial ($z$) direction, and two in the radial ($x$-$y$) plane (represented by the subscripts $+$ and $-$), with ideal characteristic frequencies \cite{Brow1986}

\begin{equation}
\nu_z = \frac{1}{2\pi} \sqrt{ \frac{qU}{md_0^2} };\,\,\,\,\,\,\, \nu_{\pm} = \frac{\nu_c}{2} \left[ 1 \pm \sqrt{ 1 - 2 \left( \frac{\nu_z}{\nu_c} \right)^2 } \right],
\label{eq::eigenfreqs_single}
\end{equation}
where $U$ is the trap depth of the electrostatic potential well, and $d_0$ a geometrical characteristic length of the trap. 
In the ideal trap (Eq.~(\ref{eq::eigenfreqs_single})), deviations in the eigenfrequencies arise from inhomogeneities and fluctuations of $U$ and $B$, linked to the geometry of the trap and the stability of the DC power supplies and of the superconducting solenoid, respectively.

We determine $\nu _c $ in this paper from a direct measurement of the three eigenfrequencies via \cite{Brow1986}
\begin{equation}
\nu _c^2=\nu_+^2+\nu_z^2+\nu_-^2. \label{eq:invariance}
\end{equation}
All the three eigenmotions can be probed individually with a dipolar electric field applied for a time $t^u_{\hbox{\scriptsize{dip}}}$ $(u=\pm,z)$ with $\nu _{\text{dip}}$ swept across an ion's eigenfrequency, analogously to a forced harmonic oscillator. Our trap configuration follows a \linebreak $\nu_{+}\gg\nu_{z}\gg\nu_{-}$~configuration, ($\nu_{+}=2.686$~MHz, $\nu_{z}=142$~kHz and $\nu_{-}=3.8$~kHz for $^{40}$Ca$^{+}$), which mitigates electric-field related uncertainties.

The experiments reported here have been carried out using a single laser-cooled ion from different calcium isotopes ($A=40,42,44,48$) in an open-ring 7-tesla Penning trap, shown schematically in Fig.~\ref{figure1}a \cite{Guti2019}. The isotope of interest is produced by photoionization in a Paul trap using two diode lasers at $423$~nm (tunable) and $375$~nm (free running), by tuning the frequency of the laser driving the $^1$S$_0\rightarrow ^1$P$_1$ transition to that of the target isotope. The ejection from the Paul trap, transport downstream the beamline, and capture into the Penning trap is as described (for $^{40}$Ca$^+$) in Ref.~\cite{Berr2022}. The voltages applied to the trap electrodes are modified after the capture to shape the trap potential with the ring electrode set to 0~V. Due to the magnetic field strength of 7~T, ten laser frequencies, tuned for each particular calcium isotope, are needed for Doppler cooling to drive the transitions shown in Fig.~\ref{figure1}b \cite{Cric2010,Guti2019}: $2\times 397$~nm for the $^2$S$_{1/2,\pm1/2}\rightarrow ^2$P$_{1/2,\pm 1/2}$ transitions (cooling),  $4\times 866$~nm for the $^2$D$_{3/2,\pm1/2,\pm 3/2}\rightarrow ^2$P$_{1/2,\pm 1/2}$ transitions (repumping) and $4\times 854$~nm for the $^2$D$_{5/2,\pm1/2,\pm 3/2}\rightarrow ^2$P$_{3/2,\pm 1/2 \pm 3/2}$ (repumping). The frequencies are distributed through twelve laser beams, two of them in the radial plane (Fig.~\ref{figure1}a). In order to speed up the cooling process, the 397-nm lasers are initially red detuned by a few GHz and scanned towards $\sim 10$~MHz below resonance, yielding a cooling time of about 30~s. During laser cooling, an axialization drive was applied to efficiently cool the magnetron motion \cite{Powe2002}. The detection of fluorescence photons from the $^2$P$_{1/2,\pm1/2}\rightarrow ^2$S$_{1/2,\pm 1/2}$ transitions (397~nm) is carried out using two devices simultaneously: an Electron-Multiplying Charge-Coupled-Device (EMCCD) and a Photo-Multiplier Tube (PMT), both at the focal plane of the $\sim 2$-m length optical system, at equal distance from a 50:50 beam splitter (BS) (Fig.~\ref{figure1}a). 

While probing any motion (applying the external dipolar field), two laser beams (dashed lines in Fig.~\ref{figure1}b) are blocked by means of an acoustic-optical modulator (stop cooling), in order to reduce the linewidth and enhance the sensitivity. Axialization is also turned off. The laser beams interact again with the ion when the dipolar electric field is stopped and axialization is again turned on. The simultaneous cooling of the radial modes implies a cooling rate for this plane lower than for the axial direction \cite{Berr2022}. The ion's radial amplitude decays exponentially with a time constant of 80(23)~ms, which allows direct observation of the excitation using the EMCCD. Figure~\ref{figure1}c shows a series of images collected within a few hundred milliseconds after probing the modified-cyclotron motion with $t^+_{\hbox{\scriptsize{dip}}}=1$~s, from which the initial amplitude ($\rho_+$) can be extracted. 
For a similar measurement of the magnetron motion $t^-_{\hbox{\scriptsize{dip}}}=100$~ms. 
The resonance curve after excitation of a radial eigenmotion is built by representing the width of the Lorentzian fitting function to the projection of the photons' distribution along the radial direction as a function of $\nu_{\hbox{\scriptsize{dip}}}$. Figure~\ref{figure1}d shows three resonance curves with different $t^+_{\hbox{\scriptsize{dip}}}$, and scaling the amplitude of the field $V^+_{\hbox{\scriptsize{dip}}}$. The axial eigenmotion had to be excited to amplitudes a factor of 3-4 larger to be sensitive with the PMT (Fig.~\ref{figure1}e). Due to the faster cooling, the excitation is not visible with the EMCCD. For the measurement in Fig.~\ref{figure1}e, $t^z_{\hbox{\scriptsize{dip}}}=100$~ms. The resonance curve after excitation is constructed from the number of detected photons (inset of Fig.~\ref{figure1}e). The amplitude of the axial motion is estimated as
\begin{equation}
\rho _z = \frac{E_z}{E_+} \frac{\nu _1}{\nu _z}\left (\rho _+ \frac{V^z_{\scriptsize{\hbox{dip}}}}{V^+_{\scriptsize{\hbox{dip}}}}\frac{t^+_{\scriptsize{\hbox{dip}}}}{2t^z_{\scriptsize{\hbox{dip}}}}+\rho _- \frac{V^z_{\scriptsize{\hbox{dip}}}}{V^-_{\scriptsize{\hbox{dip}}}}\frac{t^-_{\scriptsize{\hbox{dip}}}}{2t^z_{\scriptsize{\hbox{dip}}}}\right ), \label{eq:rho_z}
\end{equation}
where $\nu _1 =\nu _+- \nu _-$, and $E$ is the amplitude of the electric field in the center of the trap (from SIMION simulations).
\begin{figure} [t!]
\includegraphics[scale=0.50]{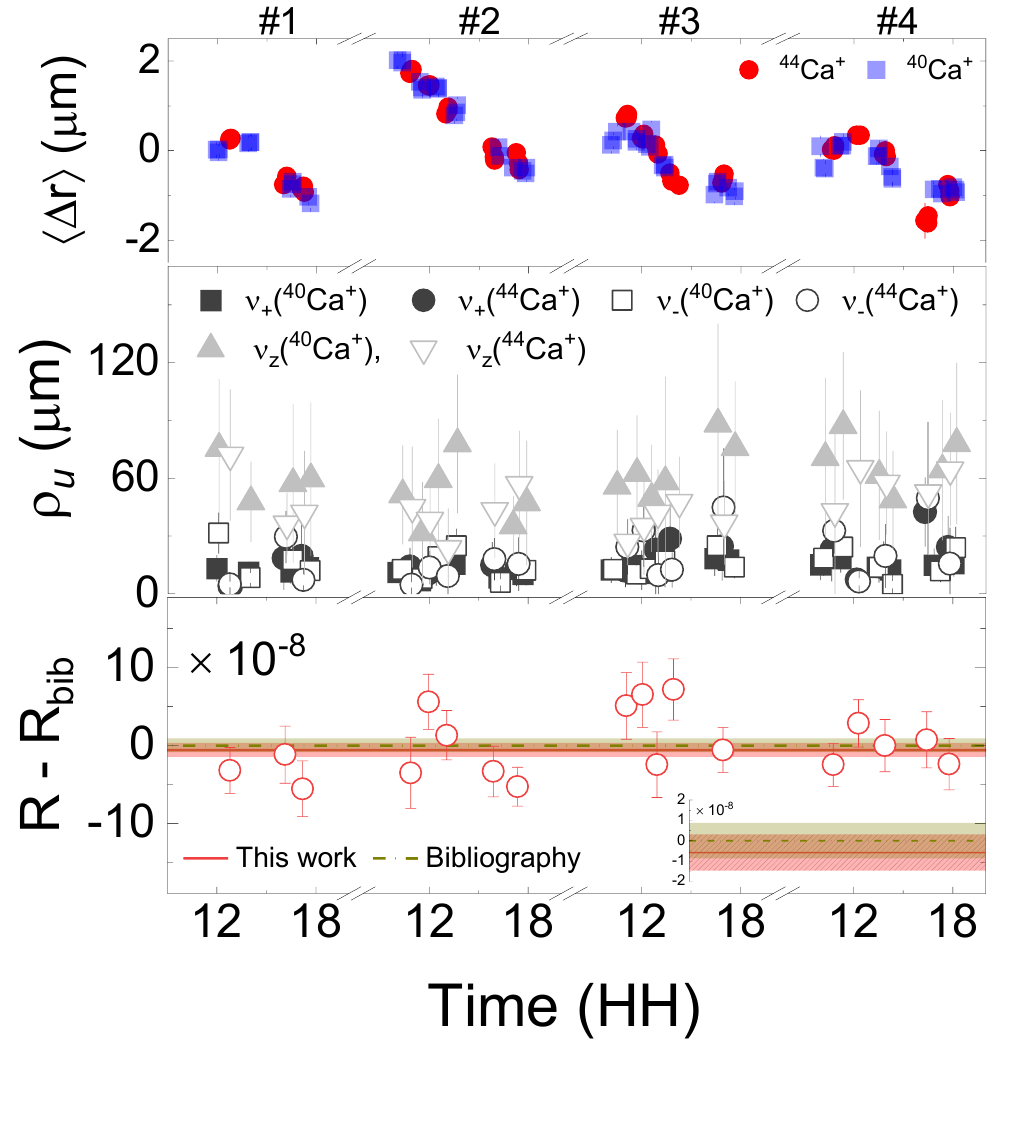}
\vspace{-14mm}
\caption{\label{figure2} Four-day measurement of the cyclotron frequencies of $^{40}$Ca$^+$ and $^{44}$Ca$^+$ including the visualization of the ion motion. Top: Deviation of the center of the ion's photon-distribution during the measurements of the radial motions. Middle: Amplitude of the modified-cyclotron ($u=+$), magnetron ($u=-$) and axial ($u=z$) motions during the measurements. Bottom: Comparison between cyclotron-frequency ratios $R=$ $\nu _c(^{40}$Ca$^+)$/$\nu _c(^{44}$Ca$^+)$ measured in this work (red-empty circles) with the values from the literature \cite{Wang2020}. The light red area (zoomed in the inset) represents the weighted mean value from this work.}
\end{figure}

We have measured sequentially the cyclotron frequency of a calcium isotope ($A=42,44,48$) and the one from $^{40}$Ca$^+$. The frequencies of the lasers in Fig.~\ref{figure1}b were accordingly tuned for each isotope, ranging from a few hundred MHz to a few GHz. Each measurement of the cyclotron-frequency ratio needs with this procedure about 1~hour, which comprises about 15~minutes per ion and 15~minutes for laser regulation. For a particular isotope, a cyclotron-frequency measurement comprises the measurement of $\nu_+$, $\nu _z$ and $\nu _ -$ (Eq.~(\ref{eq:invariance})), where the total number of scans was set to about 10/10/5, respectively. For each measurement the frequency of the external field $\nu_{\hbox{\scriptsize{dip}}}$ is swept across resonance and EMCCD images and PMT counts were simultaneously recorded. The time the rf field is applied ($t^u_{\hbox{\scriptsize{dip}}}$) and its amplitude ($V^u_{\hbox{\scriptsize{dip}}}$) are varied depending on the eigenmotion's envisaged precision. Due to the vacuum conditions and the instabilities of the DC power supplies in the experiments, $t^+_{\hbox{\scriptsize{dip}}}$ did not exceed 1~s; in most cases $t^+_{\hbox{\scriptsize{dip}}}=500$~ms. $t^z_{\hbox{\scriptsize{dip}}}$ was typically 100~ms. 

Figure~\ref{figure2} shows the cyclotron-frequency ratio \linebreak $R^{40,44}=\nu _c(^{40}$Ca$^+)/\nu _c(^{44}$Ca$^+$) from eighteen data points taken during four days of measurements. The positions and amplitudes of the ion are also shown. Our value $R^{40,44}=1.0999170772(89)$ is in agreement and with the same relative uncertainty as the most accurate value quoted in the literature $(\delta R /R)^{40,44} =8.1\times 10^{-9}$ \cite{Wang2020}.  These measurements were taken with  
$t^+_{\hbox{\scriptsize{dip}}}=500$~ms, and $t^{z,-}_{\hbox{\scriptsize{dip}}}=100$~ms. The same configuration was utilized for the measurement of the cyclotron-frequency ratio of $^{42}$Ca$^+$ - $^{40}$Ca$^+$ yielding  $R^{40,42}=1.049948070(11)$ and \linebreak $(\delta R/R) ^{40,42}=1.0\times 10^{-8}$. This result arises from eleven data points and is less precise than the one from literature $(\delta R/R)_{\scriptsize{\hbox{lit}}}^{40,42}=3.9\times 10^{-9}$. With the current performance of our system, one would need about 80~data points to reach the same level of precision. Our result of $R^{40,48}=1.199938029(15)$ with $(\delta R/R)^{40,48} =1.3\times 10^{-8}$ (based on 26 data points) is also in agreement with the quoted value $R_{\scriptsize{\hbox{lit}}}^{40,48}=1.19993802455(81)$ but with much lower precision than $(\delta R/R)_{\scriptsize{\hbox{lit}}}^{40,48} =6.8\times 10^{-10}$ \cite{Wang2020} (see Supp. material). $^{48}$Ca$^+$ has been measured with several Penning traps by means of destructive detection techniques yielding better precision than in this work \cite{Bust2013,Kwia2014,Kohl2016}. Reaching the same uncertainty level using the optical detector would demand an increase of $t^{+,z}_{\hbox{\scriptsize{dip}}}$ by a factor of $\sim 35$, and implicitly, the use of DC power supplies with higher stability. 

\begin{figure} [t!]
\includegraphics[scale=0.50]{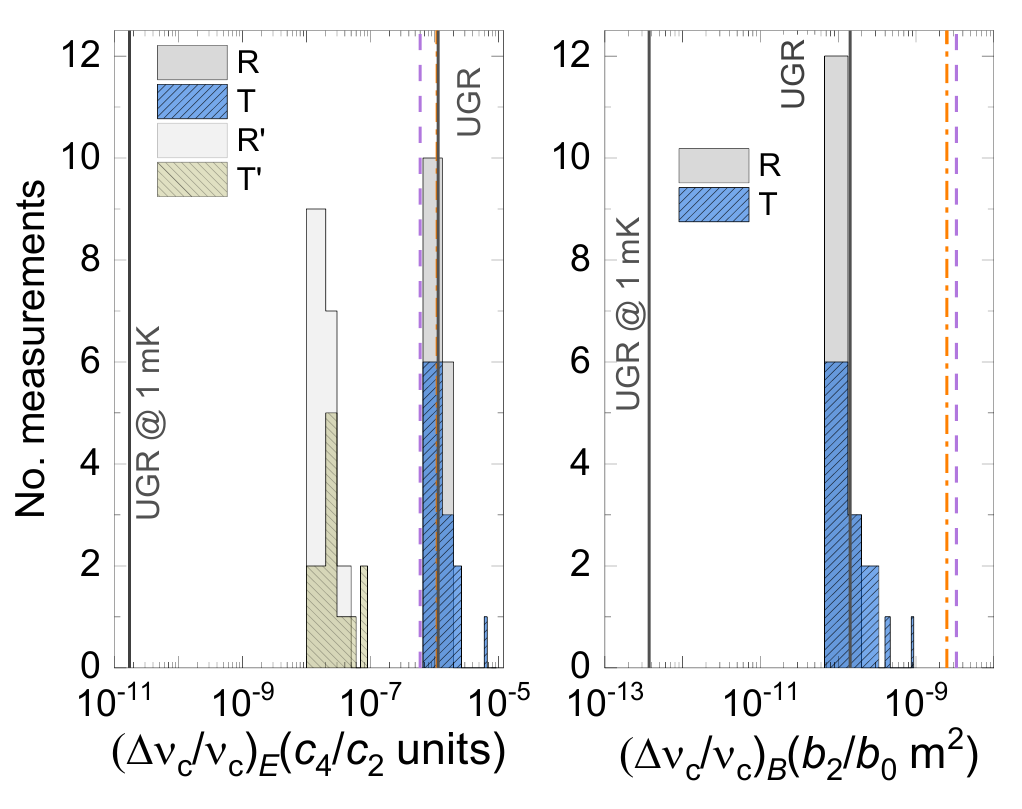}
\vspace{-8mm}
\caption{\label{figure5}Relative cyclotron-frequency shifts from the measurements on $^{44}$Ca$^+$ (T) and $^{40}$Ca$^+$ (R) as a function of $\rho _{\pm}$ (from observations) and $\rho _z$ (Eq.~\ref{eq:rho_z}), considering first-order deviations from the ideal quadrupolar field (left) and from an homogeneous magnetic field (right). The histograms are obtained from the data shown in Fig.~\ref{figure2}. The solid vertical line is the average of these measurements, the dashed line is the value assumed from experiments at Florida State University \cite{Reds2007} and the dashed dotted line the one from the ion's amplitude in experiments with highly-charged ions at the Max-Planck Institute for Nuclear Physics in Heidelberg \cite{Schu2019}, scaled to an ion with $m/q\sim 40$. The results R' and T' in the left panel are obtained considering $\rho _z$=$\rho _{\pm}$. Also the expected values when the ion's temperature is 1~mK, are shown.}
\end{figure}

Our method provides the direct observation of the radial motions, allowing for a fine tuning of amplitudes of the target and reference ion to reduce the systematic deviation due to the mass difference. Figure~\ref{figure5} shows histograms for $\Delta \nu _c/ \nu _c$ from the amplitudes of the ions' eigenmotions ($\rho _u$ in Fig.~\ref{figure2}), in units of $c_4/c_2$ and $b_2/b_0$ \cite{Kett2014}. The prime symbol stands for the same distribution but improving the sensitivity in the axial motion, by considering $\rho _z=\rho _+$. Our results are also compared with our calculations from estimated ion's oscillation amplitudes in other experiments \cite{Reds2007,Schu2019}. Note that in the case of Ref.~\cite{Schu2019}, the charge state of the ions is +17. This leads to larger eigenfrequencies, and a better value of $\Delta \nu _c/\nu _c \sim 4\times 10^{-9}$ in $c_4/c_2$ units than the one shown in Fig.~\ref{figure5} ($1.1\times 10^{-6}$), scaled for an ion with $m/q \sim 40$. 
Our uncertainty due to image charges is estimated to be $9\times 10^{-11}$ in $\nu _c$  and $2\times 10^{-11}$ in $R$ in the worst case. The uncertainty due to relativistic effects is negligible, with a maximum from our measurements of $1.2\times 10^{-13}$ (in $\nu_c$).

The use of this ion-based detector needs, in contrast to other methods, only small ion-motional amplitudes to be detected, and thus very tiny fields are required. The minimum amplitude utilized in our experiments to obtain a full resonance curve allows us extracting a force sensitivity of $50(14)$~yN$/\sqrt{\text{Hz}}$ when $\nu _{\hbox{\scriptsize{dip}}}=\nu _+$ (Fig.~\ref{figure1}d), with a signal-to-noise ratio (SNR) of 6.8(1.3). This value has been obtained from $F_r = 2m\omega _1\rho _+/t^+_{\hbox{\scriptsize{dip}}}$ with $\omega _1 =\sqrt{\omega _c ^2-\omega _z ^2}$, using the measured value of $\rho_+$ within $t^+_{\hbox{\scriptsize{dip}}}=1$~s + $t_{\hbox{\scriptsize{EMCCD}}}=75$~ms. In the axial direction $F_z=2m\omega _z\rho _z/t^z_{\hbox{\scriptsize{dip}}}$, and the sensitivity achieved for a $\text{SNR}=3.9(5)$ is $97(26)$~yN$/\sqrt{\text{Hz}}$ ($t^z_{\hbox{\scriptsize{dip}}}=100$~ms + $t_{\hbox{\scriptsize{PMT}}}=75$~ms). These values are better than previous results of $F_{z}$ in the classical regime, and are even competitive with those obtained using quantum protocols by means of Penning traps; 390(150)~yN/$\sqrt{\text{Hz}}$ at 867~kHz \cite{Bier2010} and 5.76(24)~yN/$\sqrt{\text{Hz}}$ at 1.59~MHz \cite{Gilm2021}, although our work covers a broadband spectrum and a $4\pi$ solid angle. Although an improvement in accuracy is possible as shown in Fig.~\ref{figure5}, our measurements here are presently limited by the instabilities of the DC power supplies. A long-term stability of $4.5\times 10^{-5}$ is obtained from measurements of the magnetron frequency. The magnetic field, however is stable over the level of precision in reach in this paper (See Suppl. material). Measurements with longer $t^u_{\hbox{\scriptsize{dip}}}$, will result in cyclotron frequencies with the precisions envisaged for experiments with an unbalanced Coulomb crystal ($^{40}$Ca$^+$-X$^{\scriptsize{\hbox{n}}+}$) in the same potential well of the trap, to measure the modes' eigenfrequencies of the crystal \cite{Cerr2021}.

We acknowledge support from the Spanish MICINN through the Project No. PID2019-104093GB-I00/AEI/10.01339/501100011033 and Contract No. PTA2018-016573-I, from the Andalusian Government through Project No. P18-FR-3432, from the Spanish Ministry of Education through PhD fellowship FPU17/02596, and from the University of Granada "Laboratorios Singulares 2020". The construction of the facility was supported by the European Research Council (Contract No. 278648-TRAPSENSOR), Projects No. FPA2015-67694-P and FPA2012-32076, infrastructure Projects No. UNGR10-1E-501, UNGR13-1E-1830 and EQC2018-005130-P (MICINN/FEDER/UGR), and No. IE-5713 and IE2017-5513 (Junta de Andaluc\'ia-FEDER). I. A. acknowledges support from the European Union's Horizon2020 research and innovation program under Grant Agreement No. 899354 (SuperQuLAN).


\begin{thebibliography}{36}%
\makeatletter
\providecommand \@ifxundefined [1]{%
 \@ifx{#1\undefined}
}%
\providecommand \@ifnum [1]{%
 \ifnum #1\expandafter \@firstoftwo
 \else \expandafter \@secondoftwo
 \fi
}%
\providecommand \@ifx [1]{%
 \ifx #1\expandafter \@firstoftwo
 \else \expandafter \@secondoftwo
 \fi
}%
\providecommand \natexlab [1]{#1}%
\providecommand \enquote  [1]{``#1''}%
\providecommand \bibnamefont  [1]{#1}%
\providecommand \bibfnamefont [1]{#1}%
\providecommand \citenamefont [1]{#1}%
\providecommand \href@noop [0]{\@secondoftwo}%
\providecommand \href [0]{\begingroup \@sanitize@url \@href}%
\providecommand \@href[1]{\@@startlink{#1}\@@href}%
\providecommand \@@href[1]{\endgroup#1\@@endlink}%
\providecommand \@sanitize@url [0]{\catcode `\\12\catcode `\$12\catcode
  `\&12\catcode `\#12\catcode `\^12\catcode `\_12\catcode `\%12\relax}%
\providecommand \@@startlink[1]{}%
\providecommand \@@endlink[0]{}%
\providecommand \url  [0]{\begingroup\@sanitize@url \@url }%
\providecommand \@url [1]{\endgroup\@href {#1}{\urlprefix }}%
\providecommand \urlprefix  [0]{URL }%
\providecommand \Eprint [0]{\href }%
\providecommand \doibase [0]{https://doi.org/}%
\providecommand \selectlanguage [0]{\@gobble}%
\providecommand \bibinfo  [0]{\@secondoftwo}%
\providecommand \bibfield  [0]{\@secondoftwo}%
\providecommand \translation [1]{[#1]}%
\providecommand \BibitemOpen [0]{}%
\providecommand \bibitemStop [0]{}%
\providecommand \bibitemNoStop [0]{.\EOS\space}%
\providecommand \EOS [0]{\spacefactor3000\relax}%
\providecommand \BibitemShut  [1]{\csname bibitem#1\endcsname}%
\let\auto@bib@innerbib\@empty
\bibitem [{\citenamefont {Biercuk}\ \emph {et~al.}(2010)\citenamefont
  {Biercuk}, \citenamefont {Uys}, \citenamefont {Britton}, \citenamefont
  {VanDevender},\ and\ \citenamefont {Bollinger}}]{Bier2010}%
  \BibitemOpen
  \bibfield  {author} {\bibinfo {author} {\bibfnamefont {M.~J.}\ \bibnamefont
  {Biercuk}}, \bibinfo {author} {\bibfnamefont {H.}~\bibnamefont {Uys}},
  \bibinfo {author} {\bibfnamefont {J.~W.}\ \bibnamefont {Britton}}, \bibinfo
  {author} {\bibfnamefont {A.~P.}\ \bibnamefont {VanDevender}},\ and\ \bibinfo
  {author} {\bibfnamefont {J.~J.}\ \bibnamefont {Bollinger}},\ }\bibfield
  {title} {\bibinfo {title} {Ultrasensitive detection of force and displacement
  using trapped ions},\ }\href {https://doi.org/10.1038/NNANO.2010.165}
  {\bibfield  {journal} {\bibinfo  {journal} {Nature nanotechnology}\ }\textbf
  {\bibinfo {volume} {5}},\ \bibinfo {pages} {646} (\bibinfo {year}
  {2010})}\BibitemShut {NoStop}%
\bibitem [{\citenamefont {Gilmore}\ \emph {et~al.}(2021)\citenamefont
  {Gilmore}, \citenamefont {Affolter}, \citenamefont {Lewis-Swan},
  \citenamefont {Barberena}, \citenamefont {Jordan}, \citenamefont {Rey},\ and\
  \citenamefont {Bollinger}}]{Gilm2021}%
  \BibitemOpen
  \bibfield  {author} {\bibinfo {author} {\bibfnamefont {K.~A.}\ \bibnamefont
  {Gilmore}}, \bibinfo {author} {\bibfnamefont {M.}~\bibnamefont {Affolter}},
  \bibinfo {author} {\bibfnamefont {R.~J.}\ \bibnamefont {Lewis-Swan}},
  \bibinfo {author} {\bibfnamefont {D.}~\bibnamefont {Barberena}}, \bibinfo
  {author} {\bibfnamefont {E.}~\bibnamefont {Jordan}}, \bibinfo {author}
  {\bibfnamefont {A.~M.}\ \bibnamefont {Rey}},\ and\ \bibinfo {author}
  {\bibfnamefont {J.~J.}\ \bibnamefont {Bollinger}},\ }\bibfield  {title}
  {\bibinfo {title} {Quantum-enhanced sensing of displacements and electric
  fields with two-dimensional trapped-ion crystals},\ }\href
  {https://doi.org/10.1126/science.abi5226} {\bibfield  {journal} {\bibinfo
  {journal} {Science}\ }\textbf {\bibinfo {volume} {373}},\ \bibinfo {pages}
  {673} (\bibinfo {year} {2021})}\BibitemShut {NoStop}%
\bibitem [{\citenamefont {Myers}(2019)}]{Myer2019}%
  \BibitemOpen
  \bibfield  {author} {\bibinfo {author} {\bibfnamefont {E.~G.}\ \bibnamefont
  {Myers}},\ }\bibfield  {title} {\bibinfo {title} {High-precision atomic mass
  measurements for fundamental constants},\ }\href
  {https://doi.org/10.3390/atoms7010037} {\bibfield  {journal} {\bibinfo
  {journal} {Atoms}\ }\textbf {\bibinfo {volume} {7}},\ \bibinfo {pages} {37}
  (\bibinfo {year} {2019})}\BibitemShut {NoStop}%
\bibitem [{\citenamefont {Ulmer}\ \emph {et~al.}(2015)\citenamefont {Ulmer},
  \citenamefont {Smorra}, \citenamefont {Mooser}, \citenamefont {Franke},
  \citenamefont {Nagahama}, \citenamefont {Schneider}, \citenamefont {Higuchi},
  \citenamefont {Van~Gorp}, \citenamefont {Blaum}, \citenamefont {Matsuda},
  \citenamefont {Quint}, \citenamefont {Walz},\ and\ \citenamefont
  {Yamazaki}}]{Ulme2015}%
  \BibitemOpen
  \bibfield  {author} {\bibinfo {author} {\bibfnamefont {S.}~\bibnamefont
  {Ulmer}}, \bibinfo {author} {\bibfnamefont {C.}~\bibnamefont {Smorra}},
  \bibinfo {author} {\bibfnamefont {A.}~\bibnamefont {Mooser}}, \bibinfo
  {author} {\bibfnamefont {K.}~\bibnamefont {Franke}}, \bibinfo {author}
  {\bibfnamefont {H.}~\bibnamefont {Nagahama}}, \bibinfo {author}
  {\bibfnamefont {G.}~\bibnamefont {Schneider}}, \bibinfo {author}
  {\bibfnamefont {T.}~\bibnamefont {Higuchi}}, \bibinfo {author} {\bibfnamefont
  {S.}~\bibnamefont {Van~Gorp}}, \bibinfo {author} {\bibfnamefont
  {K.}~\bibnamefont {Blaum}}, \bibinfo {author} {\bibfnamefont
  {Y.}~\bibnamefont {Matsuda}}, \bibinfo {author} {\bibfnamefont
  {W.}~\bibnamefont {Quint}}, \bibinfo {author} {\bibfnamefont
  {J.}~\bibnamefont {Walz}},\ and\ \bibinfo {author} {\bibfnamefont
  {Y.}~\bibnamefont {Yamazaki}},\ }\bibfield  {title} {\bibinfo {title}
  {High-precision comparison of the antiproton-to-proton charge-to-mass
  ratio},\ }\href {https://doi.org/10.1038/nature14861} {\bibfield  {journal}
  {\bibinfo  {journal} {Nature}\ }\textbf {\bibinfo {volume} {524}},\ \bibinfo
  {pages} {196} (\bibinfo {year} {2015})}\BibitemShut {NoStop}%
\bibitem [{\citenamefont {Myers}(2018)}]{Myer2018}%
  \BibitemOpen
  \bibfield  {author} {\bibinfo {author} {\bibfnamefont {E.~G.}\ \bibnamefont
  {Myers}},\ }\bibfield  {title} {\bibinfo {title} {{CPT} tests with the
  antihydrogen molecular ion},\ }\href
  {https://doi.org/10.1103/PhysRevA.98.010101} {\bibfield  {journal} {\bibinfo
  {journal} {Phys. Rev. A}\ }\textbf {\bibinfo {volume} {98}},\ \bibinfo
  {pages} {010101(R)} (\bibinfo {year} {2018})}\BibitemShut {NoStop}%
\bibitem [{\citenamefont {Rainville}\ \emph {et~al.}(2004)\citenamefont
  {Rainville}, \citenamefont {Thompson},\ and\ \citenamefont
  {Pritchard}}]{Rain2004}%
  \BibitemOpen
  \bibfield  {author} {\bibinfo {author} {\bibfnamefont {S.}~\bibnamefont
  {Rainville}}, \bibinfo {author} {\bibfnamefont {J.~K.}\ \bibnamefont
  {Thompson}},\ and\ \bibinfo {author} {\bibfnamefont {D.~E.}\ \bibnamefont
  {Pritchard}},\ }\bibfield  {title} {\bibinfo {title} {An ion balance for
  ultra-high-precision atomic mass measurements},\ }\href
  {https://doi.org/10.1126/science.1092320} {\bibfield  {journal} {\bibinfo
  {journal} {Science}\ }\textbf {\bibinfo {volume} {303}},\ \bibinfo {pages}
  {334} (\bibinfo {year} {2004})}\BibitemShut {NoStop}%
\bibitem [{\citenamefont {Rainville}\ \emph {et~al.}(2005)\citenamefont
  {Rainville}, \citenamefont {Thompson}, \citenamefont {Myers}, \citenamefont
  {Brown}, \citenamefont {Dewey}, \citenamefont {Kessler}, \citenamefont
  {Deslattes}, \citenamefont {B\"orner}, \citenamefont {Jentschel},
  \citenamefont {Mutti},\ and\ \citenamefont {Pritchard}}]{Rain2005}%
  \BibitemOpen
  \bibfield  {author} {\bibinfo {author} {\bibfnamefont {S.}~\bibnamefont
  {Rainville}}, \bibinfo {author} {\bibfnamefont {J.~K.}\ \bibnamefont
  {Thompson}}, \bibinfo {author} {\bibfnamefont {E.~G.}\ \bibnamefont {Myers}},
  \bibinfo {author} {\bibfnamefont {J.~M.}\ \bibnamefont {Brown}}, \bibinfo
  {author} {\bibfnamefont {M.~S.}\ \bibnamefont {Dewey}}, \bibinfo {author}
  {\bibfnamefont {E.~G.}\ \bibnamefont {Kessler}}, \bibinfo {author}
  {\bibfnamefont {R.~D.}\ \bibnamefont {Deslattes}}, \bibinfo {author}
  {\bibfnamefont {H.~G.}\ \bibnamefont {B\"orner}}, \bibinfo {author}
  {\bibfnamefont {M.}~\bibnamefont {Jentschel}}, \bibinfo {author}
  {\bibfnamefont {P.}~\bibnamefont {Mutti}},\ and\ \bibinfo {author}
  {\bibfnamefont {D.~E.}\ \bibnamefont {Pritchard}},\ }\bibfield  {title}
  {\bibinfo {title} {A direct test of $\uppercase{E}=mc^2$},\ }\href
  {https://doi.org/10.1038/4381096a} {\bibfield  {journal} {\bibinfo  {journal}
  {Nature}\ }\textbf {\bibinfo {volume} {438}},\ \bibinfo {pages} {1096}
  (\bibinfo {year} {2005})}\BibitemShut {NoStop}%
\bibitem [{\citenamefont {Myers}\ \emph {et~al.}(2015)\citenamefont {Myers},
  \citenamefont {Wagner}, \citenamefont {Kracke},\ and\ \citenamefont
  {Wenson}}]{Myer2015}%
  \BibitemOpen
  \bibfield  {author} {\bibinfo {author} {\bibfnamefont {E.~G.}\ \bibnamefont
  {Myers}}, \bibinfo {author} {\bibfnamefont {A.}~\bibnamefont {Wagner}},
  \bibinfo {author} {\bibfnamefont {H.}~\bibnamefont {Kracke}},\ and\ \bibinfo
  {author} {\bibfnamefont {B.~A.}\ \bibnamefont {Wenson}},\ }\bibfield  {title}
  {\bibinfo {title} {Atomic \uppercase{M}asses of \uppercase{T}ritium and
  \uppercase{H}elium-3},\ }\href
  {https://doi.org/10.1103/PhysRevLett.114.013003} {\bibfield  {journal}
  {\bibinfo  {journal} {Phys. Rev. Lett.}\ }\textbf {\bibinfo {volume} {114}},\
  \bibinfo {pages} {013003} (\bibinfo {year} {2015})}\BibitemShut {NoStop}%
\bibitem [{\citenamefont {Sch\"ussler}\ \emph {et~al.}(2020)\citenamefont
  {Sch\"ussler}, \citenamefont {Bekker}, \citenamefont {Brass}, \citenamefont
  {Cakir}, \citenamefont {L\'opez-Urrutia}, \citenamefont {Door}, \citenamefont
  {Filianin}, \citenamefont {Harman}, \citenamefont {Haverkort}, \citenamefont
  {Huang}, \citenamefont {Indelicato}, \citenamefont {Keitel}, \citenamefont
  {K\"onig}, \citenamefont {K.~Kromer}, \citenamefont {Novikov}, \citenamefont
  {Rischka}, \citenamefont {Schweiger}, \citenamefont {Sturm}, \citenamefont
  {Ulmer}, \citenamefont {Eliseev},\ and\ \citenamefont {Blaum}}]{Schu2020}%
  \BibitemOpen
  \bibfield  {author} {\bibinfo {author} {\bibfnamefont {R.~X.}\ \bibnamefont
  {Sch\"ussler}}, \bibinfo {author} {\bibfnamefont {H.}~\bibnamefont {Bekker}},
  \bibinfo {author} {\bibfnamefont {M.}~\bibnamefont {Brass}}, \bibinfo
  {author} {\bibfnamefont {H.}~\bibnamefont {Cakir}}, \bibinfo {author}
  {\bibfnamefont {J.~R.~C.}\ \bibnamefont {L\'opez-Urrutia}}, \bibinfo {author}
  {\bibfnamefont {M.}~\bibnamefont {Door}}, \bibinfo {author} {\bibfnamefont
  {P.}~\bibnamefont {Filianin}}, \bibinfo {author} {\bibfnamefont
  {Z.}~\bibnamefont {Harman}}, \bibinfo {author} {\bibfnamefont {M.~W.}\
  \bibnamefont {Haverkort}}, \bibinfo {author} {\bibfnamefont {W.~J.}\
  \bibnamefont {Huang}}, \bibinfo {author} {\bibfnamefont {P.}~\bibnamefont
  {Indelicato}}, \bibinfo {author} {\bibfnamefont {C.~H.}\ \bibnamefont
  {Keitel}}, \bibinfo {author} {\bibfnamefont {C.~M.}\ \bibnamefont {K\"onig}},
  \bibinfo {author} {\bibfnamefont {M.~M.}\ \bibnamefont {K.~Kromer}}, \bibinfo
  {author} {\bibfnamefont {Y.~N.}\ \bibnamefont {Novikov}}, \bibinfo {author}
  {\bibfnamefont {A.}~\bibnamefont {Rischka}}, \bibinfo {author} {\bibfnamefont
  {C.}~\bibnamefont {Schweiger}}, \bibinfo {author} {\bibfnamefont
  {S.}~\bibnamefont {Sturm}}, \bibinfo {author} {\bibfnamefont
  {S.}~\bibnamefont {Ulmer}}, \bibinfo {author} {\bibfnamefont
  {S.}~\bibnamefont {Eliseev}},\ and\ \bibinfo {author} {\bibfnamefont
  {K.}~\bibnamefont {Blaum}},\ }\bibfield  {title} {\bibinfo {title} {Detection
  of metastable electronic states by {P}enning trap mass spectrometry},\ }\href
  {https://doi.org/10.1038/s41586-020-2221-0} {\bibfield  {journal} {\bibinfo
  {journal} {Nature}\ }\textbf {\bibinfo {volume} {581}},\ \bibinfo {pages}
  {42} (\bibinfo {year} {2020})}\BibitemShut {NoStop}%
\bibitem [{\citenamefont {Wineland}\ and\ \citenamefont
  {Dehmelt}(1975)}]{Wine1975}%
  \BibitemOpen
  \bibfield  {author} {\bibinfo {author} {\bibfnamefont {D.~J.}\ \bibnamefont
  {Wineland}}\ and\ \bibinfo {author} {\bibfnamefont {H.~G.}\ \bibnamefont
  {Dehmelt}},\ }\bibfield  {title} {\bibinfo {title} {Principles of the stored
  ion calorimeter},\ }\href {https://doi.org/10.1063/1.321602} {\bibfield
  {journal} {\bibinfo  {journal} {J.\ Appl.\ Phys.}\ }\textbf {\bibinfo
  {volume} {46}},\ \bibinfo {pages} {919} (\bibinfo {year} {1975})}\BibitemShut
  {NoStop}%
\bibitem [{\citenamefont {Brown}\ and\ \citenamefont
  {Gabrielse}(1986)}]{Brow1986}%
  \BibitemOpen
  \bibfield  {author} {\bibinfo {author} {\bibfnamefont {L.~S.}\ \bibnamefont
  {Brown}}\ and\ \bibinfo {author} {\bibfnamefont {G.}~\bibnamefont
  {Gabrielse}},\ }\bibfield  {title} {\bibinfo {title} {Geonium theory: Physics
  of a single electron or ion in a \uppercase{P}enning trap},\ }\href
  {https://doi.org/10.1103/RevModPhys.58.233} {\bibfield  {journal} {\bibinfo
  {journal} {Rev. Mod. Phys.}\ }\textbf {\bibinfo {volume} {58}},\ \bibinfo
  {pages} {233} (\bibinfo {year} {1986})}\BibitemShut {NoStop}%
\bibitem [{\citenamefont {Filianin}\ \emph {et~al.}(2021)\citenamefont
  {Filianin}, \citenamefont {Lyu}, \citenamefont {Door}, \citenamefont {Blaum},
  \citenamefont {Huang}, \citenamefont {Haverkort}, \citenamefont {Indelicato},
  \citenamefont {Keitel}, \citenamefont {Kromer}, \citenamefont {Lange},
  \citenamefont {Novikov}, \citenamefont {Rischka}, \citenamefont
  {Sch\"ussler}, \citenamefont {Schweiger}, \citenamefont {Sturm},
  \citenamefont {Ulmer}, \citenamefont {Harman},\ and\ \citenamefont
  {Eliseev}}]{Fili2021}%
  \BibitemOpen
  \bibfield  {author} {\bibinfo {author} {\bibfnamefont {P.}~\bibnamefont
  {Filianin}}, \bibinfo {author} {\bibfnamefont {C.}~\bibnamefont {Lyu}},
  \bibinfo {author} {\bibfnamefont {M.}~\bibnamefont {Door}}, \bibinfo {author}
  {\bibfnamefont {K.}~\bibnamefont {Blaum}}, \bibinfo {author} {\bibfnamefont
  {W.~J.}\ \bibnamefont {Huang}}, \bibinfo {author} {\bibfnamefont
  {M.}~\bibnamefont {Haverkort}}, \bibinfo {author} {\bibfnamefont
  {P.}~\bibnamefont {Indelicato}}, \bibinfo {author} {\bibfnamefont {C.~H.}\
  \bibnamefont {Keitel}}, \bibinfo {author} {\bibfnamefont {K.}~\bibnamefont
  {Kromer}}, \bibinfo {author} {\bibfnamefont {D.}~\bibnamefont {Lange}},
  \bibinfo {author} {\bibfnamefont {Y.~N.}\ \bibnamefont {Novikov}}, \bibinfo
  {author} {\bibfnamefont {A.}~\bibnamefont {Rischka}}, \bibinfo {author}
  {\bibfnamefont {R.~X.}\ \bibnamefont {Sch\"ussler}}, \bibinfo {author}
  {\bibfnamefont {C.}~\bibnamefont {Schweiger}}, \bibinfo {author}
  {\bibfnamefont {S.}~\bibnamefont {Sturm}}, \bibinfo {author} {\bibfnamefont
  {S.}~\bibnamefont {Ulmer}}, \bibinfo {author} {\bibfnamefont
  {Z.}~\bibnamefont {Harman}},\ and\ \bibinfo {author} {\bibfnamefont
  {S.}~\bibnamefont {Eliseev}},\ }\bibfield  {title} {\bibinfo {title} {Direct
  $\uppercase{Q}$-value determination of the $\beta ^-$ decay of
  $^{187}$\uppercase{R}e},\ }\href
  {https://doi.org/10.1103/PhysRevLett.127.072502} {\bibfield  {journal}
  {\bibinfo  {journal} {Phys. Rev. Lett.}\ }\textbf {\bibinfo {volume} {127}},\
  \bibinfo {pages} {072502} (\bibinfo {year} {2021})}\BibitemShut {NoStop}%
\bibitem [{\citenamefont {Weisskoff}\ \emph {et~al.}(1988)\citenamefont
  {Weisskoff}, \citenamefont {Lafyatis}, \citenamefont {Boyce}, \citenamefont
  {Cornell}, \citenamefont {Flanagan},\ and\ \citenamefont
  {Pritchard}}]{Weis1988}%
  \BibitemOpen
  \bibfield  {author} {\bibinfo {author} {\bibfnamefont {R.~M.}\ \bibnamefont
  {Weisskoff}}, \bibinfo {author} {\bibfnamefont {G.~P.}\ \bibnamefont
  {Lafyatis}}, \bibinfo {author} {\bibfnamefont {K.~R.}\ \bibnamefont {Boyce}},
  \bibinfo {author} {\bibfnamefont {E.~A.}\ \bibnamefont {Cornell}}, \bibinfo
  {author} {\bibfnamefont {R.~W.}\ \bibnamefont {Flanagan}},\ and\ \bibinfo
  {author} {\bibfnamefont {D.}~\bibnamefont {Pritchard}},\ }\bibfield  {title}
  {\bibinfo {title} {rf \uppercase{SQUID} detector for single‐ion trapping
  experiments},\ }\href {https://doi.org/10.1063/1.340137} {\bibfield
  {journal} {\bibinfo  {journal} {J.\ Appl.\ Phys.}\ }\textbf {\bibinfo
  {volume} {63}},\ \bibinfo {pages} {4599} (\bibinfo {year}
  {1988})}\BibitemShut {NoStop}%
\bibitem [{\citenamefont {Ulmer}\ \emph {et~al.}(2009)\citenamefont {Ulmer},
  \citenamefont {Kracke}, \citenamefont {Blaum}, \citenamefont {Kreim},
  \citenamefont {Mooser}, \citenamefont {Quint}, \citenamefont {Rodegheri},\
  and\ \citenamefont {Walz}}]{Ulme2009}%
  \BibitemOpen
  \bibfield  {author} {\bibinfo {author} {\bibfnamefont {S.}~\bibnamefont
  {Ulmer}}, \bibinfo {author} {\bibfnamefont {H.}~\bibnamefont {Kracke}},
  \bibinfo {author} {\bibfnamefont {K.}~\bibnamefont {Blaum}}, \bibinfo
  {author} {\bibfnamefont {S.}~\bibnamefont {Kreim}}, \bibinfo {author}
  {\bibfnamefont {A.}~\bibnamefont {Mooser}}, \bibinfo {author} {\bibfnamefont
  {W.}~\bibnamefont {Quint}}, \bibinfo {author} {\bibfnamefont {C.~C.}\
  \bibnamefont {Rodegheri}},\ and\ \bibinfo {author} {\bibfnamefont
  {J.}~\bibnamefont {Walz}},\ }\bibfield  {title} {\bibinfo {title} {The
  quality factor of a superconducting rf resonator in a magnetic field},\
  }\href {https://doi.org/10.1063/1.3271537} {\bibfield  {journal} {\bibinfo
  {journal} {Rev.\ Sci.\ Instrum.}\ }\textbf {\bibinfo {volume} {80}},\
  \bibinfo {pages} {123302} (\bibinfo {year} {2009})}\BibitemShut {NoStop}%
\bibitem [{\citenamefont {Cornell}\ \emph {et~al.}(1990)\citenamefont
  {Cornell}, \citenamefont {Weisskoff}, \citenamefont {Boyce},\ and\
  \citenamefont {Pritchard}}]{Corn1990}%
  \BibitemOpen
  \bibfield  {author} {\bibinfo {author} {\bibfnamefont {E.~A.}\ \bibnamefont
  {Cornell}}, \bibinfo {author} {\bibfnamefont {R.~M.}\ \bibnamefont
  {Weisskoff}}, \bibinfo {author} {\bibfnamefont {K.~R.}\ \bibnamefont
  {Boyce}},\ and\ \bibinfo {author} {\bibfnamefont {D.~E.}\ \bibnamefont
  {Pritchard}},\ }\bibfield  {title} {\bibinfo {title} {Mode coupling in a
  {P}enning trap: $\pi $-pulses and a classical avoided crossing},\ }\href
  {https://doi.org/10.1103/PhysRevA.41.312} {\bibfield  {journal} {\bibinfo
  {journal} {Phys. Rev. A}\ }\textbf {\bibinfo {volume} {41}},\ \bibinfo
  {pages} {312} (\bibinfo {year} {1990})}\BibitemShut {NoStop}%
\bibitem [{\citenamefont {Wineland}\ \emph {et~al.}(1983)\citenamefont
  {Wineland}, \citenamefont {Bollinger},\ and\ \citenamefont
  {Itano}}]{Wine1983}%
  \BibitemOpen
  \bibfield  {author} {\bibinfo {author} {\bibfnamefont {D.~J.}\ \bibnamefont
  {Wineland}}, \bibinfo {author} {\bibfnamefont {J.~J.}\ \bibnamefont
  {Bollinger}},\ and\ \bibinfo {author} {\bibfnamefont {W.~M.}\ \bibnamefont
  {Itano}},\ }\bibfield  {title} {\bibinfo {title} {Laser-fluorescence mass
  spectroscopy},\ }\href {https://doi.org/10.1103/PhysRevLett.50.628}
  {\bibfield  {journal} {\bibinfo  {journal} {Phys. Rev. Lett.}\ }\textbf
  {\bibinfo {volume} {50}},\ \bibinfo {pages} {628} (\bibinfo {year}
  {1983})}\BibitemShut {NoStop}%
\bibitem [{\citenamefont {Nagy}\ \emph {et~al.}(2006)\citenamefont {Nagy},
  \citenamefont {Fritioff}, \citenamefont {Solders}, \citenamefont {Schuch},
  \citenamefont {Bj\"orkhage},\ and\ \citenamefont {Bergstr\"om}}]{Nagy2006}%
  \BibitemOpen
  \bibfield  {author} {\bibinfo {author} {\bibfnamefont {S.}~\bibnamefont
  {Nagy}}, \bibinfo {author} {\bibfnamefont {T.}~\bibnamefont {Fritioff}},
  \bibinfo {author} {\bibfnamefont {A.}~\bibnamefont {Solders}}, \bibinfo
  {author} {\bibfnamefont {R.}~\bibnamefont {Schuch}}, \bibinfo {author}
  {\bibfnamefont {M.}~\bibnamefont {Bj\"orkhage}},\ and\ \bibinfo {author}
  {\bibfnamefont {I.}~\bibnamefont {Bergstr\"om}},\ }\bibfield  {title}
  {\bibinfo {title} {Precision mass measurements of
  $^{40}$\uppercase{C}a$^{17+}$ and $^{40}$\uppercase{C}a$^{19+}$ ions in a
  \uppercase{P}enning trap},\ }\href
  {https://doi.org/10.1140/epjd/e2006-00064-8} {\bibfield  {journal} {\bibinfo
  {journal} {Eur. Phys. J. D}\ }\textbf {\bibinfo {volume} {39}},\ \bibinfo
  {pages} {1} (\bibinfo {year} {2006})}\BibitemShut {NoStop}%
\bibitem [{\citenamefont {Gebert}\ \emph {et~al.}(2015)\citenamefont {Gebert},
  \citenamefont {Wan}, \citenamefont {Wolf}, \citenamefont {Angstmann},
  \citenamefont {Berengut},\ and\ \citenamefont {Schmidt}}]{Gebe2015}%
  \BibitemOpen
  \bibfield  {author} {\bibinfo {author} {\bibfnamefont {F.}~\bibnamefont
  {Gebert}}, \bibinfo {author} {\bibfnamefont {Y.}~\bibnamefont {Wan}},
  \bibinfo {author} {\bibfnamefont {F.}~\bibnamefont {Wolf}}, \bibinfo {author}
  {\bibfnamefont {C.~N.}\ \bibnamefont {Angstmann}}, \bibinfo {author}
  {\bibfnamefont {J.~C.}\ \bibnamefont {Berengut}},\ and\ \bibinfo {author}
  {\bibfnamefont {P.~O.}\ \bibnamefont {Schmidt}},\ }\bibfield  {title}
  {\bibinfo {title} {Precision isotope shift measurements in calcium ions using
  quantum logic detection schemes},\ }\href
  {https://doi.org/10.1103/PhysRevLett.115.053003} {\bibfield  {journal}
  {\bibinfo  {journal} {Phys.\ Rev.\ Lett.}\ }\textbf {\bibinfo {volume}
  {115}},\ \bibinfo {pages} {053003} (\bibinfo {year} {2015})}\BibitemShut
  {NoStop}%
\bibitem [{\citenamefont {Knollmann}\ \emph {et~al.}(2019)\citenamefont
  {Knollmann}, \citenamefont {Patel},\ and\ \citenamefont {Doret}}]{Knol2019}%
  \BibitemOpen
  \bibfield  {author} {\bibinfo {author} {\bibfnamefont {F.}~\bibnamefont
  {Knollmann}}, \bibinfo {author} {\bibfnamefont {A.}~\bibnamefont {Patel}},\
  and\ \bibinfo {author} {\bibfnamefont {C.}~\bibnamefont {Doret}},\ }\bibfield
   {title} {\bibinfo {title} {Part-per-billion meaurement of the
  4$^2\uppercase{S}_{1/2}\rightarrow 3^2\uppercase{D}_{5/2}$
  electric-quadrupole transition isotope shifts between
  $^{42,44,48}$\uppercase{C}a$^+$ and $^{40}$\uppercase{C}a$^+$},\ }\href
  {https://doi.org/10.1103/PhysRevA.100.022514} {\bibfield  {journal} {\bibinfo
   {journal} {Phys. Rev. A}\ }\textbf {\bibinfo {volume} {100}},\ \bibinfo
  {pages} {022514} (\bibinfo {year} {2019})}\BibitemShut {NoStop}%
\bibitem [{\citenamefont {Solaro}\ \emph {et~al.}(2020)\citenamefont {Solaro},
  \citenamefont {Meyer}, \citenamefont {Fisher}, \citenamefont {Berengut},
  \citenamefont {Fuchs},\ and\ \citenamefont {Drewsen}}]{Sola2020}%
  \BibitemOpen
  \bibfield  {author} {\bibinfo {author} {\bibfnamefont {C.}~\bibnamefont
  {Solaro}}, \bibinfo {author} {\bibfnamefont {S.}~\bibnamefont {Meyer}},
  \bibinfo {author} {\bibfnamefont {K.}~\bibnamefont {Fisher}}, \bibinfo
  {author} {\bibfnamefont {J.}~\bibnamefont {Berengut}}, \bibinfo {author}
  {\bibfnamefont {E.}~\bibnamefont {Fuchs}},\ and\ \bibinfo {author}
  {\bibfnamefont {M.}~\bibnamefont {Drewsen}},\ }\bibfield  {title} {\bibinfo
  {title} {Improved isotope-shift-based bounds on bosons beyond the standard
  model through measurements of the $^{2}$\uppercase{D}$_{3/2}$-
  $^{2}$\uppercase{D}$_{5/2}$ interval in \uppercase{C}a$^+$},\ }\href
  {https://doi.org/10.1103/PhysRevLett.125.123003} {\bibfield  {journal}
  {\bibinfo  {journal} {Phys. Rev. Lett.}\ }\textbf {\bibinfo {volume} {125}},\
  \bibinfo {pages} {123003} (\bibinfo {year} {2020})}\BibitemShut {NoStop}%
\bibitem [{\citenamefont {Guti{\'e}rrez}\ \emph
  {et~al.}(2019{\natexlab{a}})\citenamefont {Guti{\'e}rrez}, \citenamefont
  {Berrocal}, \citenamefont {Dom{\'i}nguez}, \citenamefont {Arrazola},
  \citenamefont {Block}, \citenamefont {Solano},\ and\ \citenamefont
  {Rodr{\'i}guez}}]{Guti2019b}%
  \BibitemOpen
  \bibfield  {author} {\bibinfo {author} {\bibfnamefont {M.~J.}\ \bibnamefont
  {Guti{\'e}rrez}}, \bibinfo {author} {\bibfnamefont {J.}~\bibnamefont
  {Berrocal}}, \bibinfo {author} {\bibfnamefont {F.}~\bibnamefont
  {Dom{\'i}nguez}}, \bibinfo {author} {\bibfnamefont {I.}~\bibnamefont
  {Arrazola}}, \bibinfo {author} {\bibfnamefont {M.}~\bibnamefont {Block}},
  \bibinfo {author} {\bibfnamefont {E.}~\bibnamefont {Solano}},\ and\ \bibinfo
  {author} {\bibfnamefont {D.}~\bibnamefont {Rodr{\'i}guez}},\ }\bibfield
  {title} {\bibinfo {title} {Dynamics of an unbalanced two-ion crystal in a
  {P}enning trap for application in optical mass spectrometry},\ }\href
  {https://doi.org/10.1103/PhysRevA.100.063415} {\bibfield  {journal} {\bibinfo
   {journal} {Phys. Rev. A}\ }\textbf {\bibinfo {volume} {100}},\ \bibinfo
  {pages} {063415} (\bibinfo {year} {2019}{\natexlab{a}})}\BibitemShut
  {NoStop}%
\bibitem [{\citenamefont {Cerrillo}\ and\ \citenamefont
  {Rodr\'iguez}(2021)}]{Cerr2021}%
  \BibitemOpen
  \bibfield  {author} {\bibinfo {author} {\bibfnamefont {J.}~\bibnamefont
  {Cerrillo}}\ and\ \bibinfo {author} {\bibfnamefont {D.}~\bibnamefont
  {Rodr\'iguez}},\ }\bibfield  {title} {\bibinfo {title} {Motional quantum
  metrology in a {P}enning trap},\ }\href
  {https://doi.org/10.1209/0295-5075/134/38001} {\bibfield  {journal} {\bibinfo
   {journal} {EPL}\ }\textbf {\bibinfo {volume} {134}},\ \bibinfo {pages}
  {38001} (\bibinfo {year} {2021})}\BibitemShut {NoStop}%
\bibitem [{\citenamefont {Block}\ \emph {et~al.}(2010)\citenamefont {Block},
  \citenamefont {Ackermann}, \citenamefont {Blaum}, \citenamefont {Droese},
  \citenamefont {Dworschak}, \citenamefont {Eliseev}, \citenamefont
  {Fleckenstein}, \citenamefont {Haettner}, \citenamefont {Herfurth},
  \citenamefont {He{\ss}berger}, \citenamefont {Hofmann}, \citenamefont
  {Ketelaer}, \citenamefont {Ketter}, \citenamefont {Kluge}, \citenamefont
  {Marx}, \citenamefont {Mazzocco}, \citenamefont {Novikov}, \citenamefont
  {Pla{\ss}}, \citenamefont {Popeko}, \citenamefont {Rahaman}, \citenamefont
  {{Rodr{\'i}guez}}, \citenamefont {Scheidenberger}, \citenamefont
  {Schweikhard}, \citenamefont {Thirolf}, \citenamefont {Vorobyev},\ and\
  \citenamefont {Weber}}]{Bloc2010}%
  \BibitemOpen
  \bibfield  {author} {\bibinfo {author} {\bibfnamefont {M.}~\bibnamefont
  {Block}}, \bibinfo {author} {\bibfnamefont {D.}~\bibnamefont {Ackermann}},
  \bibinfo {author} {\bibfnamefont {K.}~\bibnamefont {Blaum}}, \bibinfo
  {author} {\bibfnamefont {C.}~\bibnamefont {Droese}}, \bibinfo {author}
  {\bibfnamefont {M.}~\bibnamefont {Dworschak}}, \bibinfo {author}
  {\bibfnamefont {S.}~\bibnamefont {Eliseev}}, \bibinfo {author} {\bibfnamefont
  {T.}~\bibnamefont {Fleckenstein}}, \bibinfo {author} {\bibfnamefont
  {E.}~\bibnamefont {Haettner}}, \bibinfo {author} {\bibfnamefont
  {F.}~\bibnamefont {Herfurth}}, \bibinfo {author} {\bibfnamefont {F.~P.}\
  \bibnamefont {He{\ss}berger}}, \bibinfo {author} {\bibfnamefont
  {S.}~\bibnamefont {Hofmann}}, \bibinfo {author} {\bibfnamefont
  {J.}~\bibnamefont {Ketelaer}}, \bibinfo {author} {\bibfnamefont
  {J.}~\bibnamefont {Ketter}}, \bibinfo {author} {\bibfnamefont {H.-J.}\
  \bibnamefont {Kluge}}, \bibinfo {author} {\bibfnamefont {G.}~\bibnamefont
  {Marx}}, \bibinfo {author} {\bibfnamefont {M.}~\bibnamefont {Mazzocco}},
  \bibinfo {author} {\bibfnamefont {Y.~N.}\ \bibnamefont {Novikov}}, \bibinfo
  {author} {\bibfnamefont {W.~R.}\ \bibnamefont {Pla{\ss}}}, \bibinfo {author}
  {\bibfnamefont {A.}~\bibnamefont {Popeko}}, \bibinfo {author} {\bibfnamefont
  {S.}~\bibnamefont {Rahaman}}, \bibinfo {author} {\bibfnamefont
  {D.}~\bibnamefont {{Rodr{\'i}guez}}}, \bibinfo {author} {\bibfnamefont
  {C.}~\bibnamefont {Scheidenberger}}, \bibinfo {author} {\bibfnamefont
  {L.}~\bibnamefont {Schweikhard}}, \bibinfo {author} {\bibfnamefont {P.~G.}\
  \bibnamefont {Thirolf}}, \bibinfo {author} {\bibfnamefont {G.~K.}\
  \bibnamefont {Vorobyev}},\ and\ \bibinfo {author} {\bibfnamefont
  {C.}~\bibnamefont {Weber}},\ }\bibfield  {title} {\bibinfo {title} {Direct
  mass measurements above uranium bridge the gap to the island of stability},\
  }\href {https://doi.org/10.1038/nature08774} {\bibfield  {journal} {\bibinfo
  {journal} {Nature}\ }\textbf {\bibinfo {volume} {463}},\ \bibinfo {pages}
  {785} (\bibinfo {year} {2010})}\BibitemShut {NoStop}%
\bibitem [{\citenamefont {Minaya~Ramirez}\ \emph {et~al.}(2012)\citenamefont
  {Minaya~Ramirez}, \citenamefont {Ackermann}, \citenamefont {Blaum},
  \citenamefont {Block}, \citenamefont {Droese}, \citenamefont
  {D{\"{u}}llmann}, \citenamefont {Dworschak}, \citenamefont {Eibach},
  \citenamefont {Eliseev}, \citenamefont {Haettner}, \citenamefont {Herfurth},
  \citenamefont {He{\ss}berger}, \citenamefont {Hofmann}, \citenamefont
  {Ketelaer}, \citenamefont {Marx}, \citenamefont {Mazzocco}, \citenamefont
  {Nesterenko}, \citenamefont {Novikov}, \citenamefont {Pla{\ss}},
  \citenamefont {Rodr{\'\i}guez}, \citenamefont {Scheidenberger}, \citenamefont
  {Schweikhard}, \citenamefont {Thirolf},\ and\ \citenamefont
  {Weber}}]{Mina2012}%
  \BibitemOpen
  \bibfield  {author} {\bibinfo {author} {\bibfnamefont {E.}~\bibnamefont
  {Minaya~Ramirez}}, \bibinfo {author} {\bibfnamefont {D.}~\bibnamefont
  {Ackermann}}, \bibinfo {author} {\bibfnamefont {K.}~\bibnamefont {Blaum}},
  \bibinfo {author} {\bibfnamefont {M.}~\bibnamefont {Block}}, \bibinfo
  {author} {\bibfnamefont {C.}~\bibnamefont {Droese}}, \bibinfo {author}
  {\bibfnamefont {{\relax Ch}.~E.}\ \bibnamefont {D{\"{u}}llmann}}, \bibinfo
  {author} {\bibfnamefont {M.}~\bibnamefont {Dworschak}}, \bibinfo {author}
  {\bibfnamefont {M.}~\bibnamefont {Eibach}}, \bibinfo {author} {\bibfnamefont
  {S.}~\bibnamefont {Eliseev}}, \bibinfo {author} {\bibfnamefont
  {E.}~\bibnamefont {Haettner}}, \bibinfo {author} {\bibfnamefont
  {F.}~\bibnamefont {Herfurth}}, \bibinfo {author} {\bibfnamefont {F.~P.}\
  \bibnamefont {He{\ss}berger}}, \bibinfo {author} {\bibfnamefont
  {S.}~\bibnamefont {Hofmann}}, \bibinfo {author} {\bibfnamefont
  {J.}~\bibnamefont {Ketelaer}}, \bibinfo {author} {\bibfnamefont
  {G.}~\bibnamefont {Marx}}, \bibinfo {author} {\bibfnamefont {M.}~\bibnamefont
  {Mazzocco}}, \bibinfo {author} {\bibfnamefont {D.}~\bibnamefont
  {Nesterenko}}, \bibinfo {author} {\bibfnamefont {{\relax Yu}.~N.}\
  \bibnamefont {Novikov}}, \bibinfo {author} {\bibfnamefont {W.~R.}\
  \bibnamefont {Pla{\ss}}}, \bibinfo {author} {\bibfnamefont {D.}~\bibnamefont
  {Rodr{\'\i}guez}}, \bibinfo {author} {\bibfnamefont {C.}~\bibnamefont
  {Scheidenberger}}, \bibinfo {author} {\bibfnamefont {L.}~\bibnamefont
  {Schweikhard}}, \bibinfo {author} {\bibfnamefont {P.~G.}\ \bibnamefont
  {Thirolf}},\ and\ \bibinfo {author} {\bibfnamefont {C.}~\bibnamefont
  {Weber}},\ }\bibfield  {title} {\bibinfo {title} {Direct mapping of nuclear
  shell effects in the heaviest elements},\ }\href
  {https://doi.org/10.1126/science.1225636} {\bibfield  {journal} {\bibinfo
  {journal} {Science}\ }\textbf {\bibinfo {volume} {337}},\ \bibinfo {pages}
  {1207} (\bibinfo {year} {2012})}\BibitemShut {NoStop}%
\bibitem [{\citenamefont {Kaleja}\ \emph {et~al.}(2022)\citenamefont {Kaleja},
  \citenamefont {Andeli\'c}, \citenamefont {Bezrodnova}, \citenamefont {Blaum},
  \citenamefont {Block}, \citenamefont {Chenmarev}, \citenamefont {Chhetri},
  \citenamefont {D\"ullmann}, \citenamefont {Eibach}, \citenamefont {Eliseev},
  \citenamefont {Even}, \citenamefont {Filianin}, \citenamefont {Giacoppo},
  \citenamefont {G\"otz}, \citenamefont {Gusev}, \citenamefont {Guti\'errez},
  \citenamefont {Hessberger}, \citenamefont {Kalantar-Nayestanaki},
  \citenamefont {van~de Laar}, \citenamefont {Laatiaoui}, \citenamefont
  {Lohse}, \citenamefont {Martynova}, \citenamefont {Ramirez}, \citenamefont
  {Mistry}, \citenamefont {Murb\"ock}, \citenamefont {Novikov}, \citenamefont
  {Raeder}, \citenamefont {Rodr\'iguez}, \citenamefont {Schneider},
  \citenamefont {Schweikhard}, \citenamefont {Thirolf1},\ and\ \citenamefont
  {Yakushev}}]{Kale2022}%
  \BibitemOpen
  \bibfield  {author} {\bibinfo {author} {\bibfnamefont {O.}~\bibnamefont
  {Kaleja}}, \bibinfo {author} {\bibfnamefont {B.}~\bibnamefont {Andeli\'c}},
  \bibinfo {author} {\bibfnamefont {O.}~\bibnamefont {Bezrodnova}}, \bibinfo
  {author} {\bibfnamefont {K.}~\bibnamefont {Blaum}}, \bibinfo {author}
  {\bibfnamefont {M.}~\bibnamefont {Block}}, \bibinfo {author} {\bibfnamefont
  {S.}~\bibnamefont {Chenmarev}}, \bibinfo {author} {\bibfnamefont
  {P.}~\bibnamefont {Chhetri}}, \bibinfo {author} {\bibfnamefont {C.~D. C.~E.}\
  \bibnamefont {D\"ullmann}}, \bibinfo {author} {\bibfnamefont
  {M.}~\bibnamefont {Eibach}}, \bibinfo {author} {\bibfnamefont
  {S.}~\bibnamefont {Eliseev}}, \bibinfo {author} {\bibfnamefont
  {J.}~\bibnamefont {Even}}, \bibinfo {author} {\bibfnamefont {P.}~\bibnamefont
  {Filianin}}, \bibinfo {author} {\bibfnamefont {F.}~\bibnamefont {Giacoppo}},
  \bibinfo {author} {\bibfnamefont {S.}~\bibnamefont {G\"otz}}, \bibinfo
  {author} {\bibfnamefont {Y.}~\bibnamefont {Gusev}}, \bibinfo {author}
  {\bibfnamefont {M.~J.}\ \bibnamefont {Guti\'errez}}, \bibinfo {author}
  {\bibfnamefont {F.~P.}\ \bibnamefont {Hessberger}}, \bibinfo {author}
  {\bibfnamefont {N.}~\bibnamefont {Kalantar-Nayestanaki}}, \bibinfo {author}
  {\bibfnamefont {J.~J.~W.}\ \bibnamefont {van~de Laar}}, \bibinfo {author}
  {\bibfnamefont {M.}~\bibnamefont {Laatiaoui}}, \bibinfo {author}
  {\bibfnamefont {S.}~\bibnamefont {Lohse}}, \bibinfo {author} {\bibfnamefont
  {N.}~\bibnamefont {Martynova}}, \bibinfo {author} {\bibfnamefont {E.~M.}\
  \bibnamefont {Ramirez}}, \bibinfo {author} {\bibfnamefont {A.~K.}\
  \bibnamefont {Mistry}}, \bibinfo {author} {\bibfnamefont {T.}~\bibnamefont
  {Murb\"ock}}, \bibinfo {author} {\bibfnamefont {Y.}~\bibnamefont {Novikov}},
  \bibinfo {author} {\bibfnamefont {S.}~\bibnamefont {Raeder}}, \bibinfo
  {author} {\bibfnamefont {D.}~\bibnamefont {Rodr\'iguez}}, \bibinfo {author}
  {\bibfnamefont {F.}~\bibnamefont {Schneider}}, \bibinfo {author}
  {\bibfnamefont {L.}~\bibnamefont {Schweikhard}}, \bibinfo {author}
  {\bibfnamefont {P.~G.}\ \bibnamefont {Thirolf1}},\ and\ \bibinfo {author}
  {\bibfnamefont {A.}~\bibnamefont {Yakushev}},\ }\bibfield  {title} {\bibinfo
  {title} {Direct high-precision mass spectrometry of superheavy elements with
  \uppercase{SHIPTRAP}},\ }\href {https://doi.org/10.1103/PhysRevC.106.054325}
  {\bibfield  {journal} {\bibinfo  {journal} {Phys. Rev. C}\ }\textbf {\bibinfo
  {volume} {106}},\ \bibinfo {pages} {054325} (\bibinfo {year}
  {2022})}\BibitemShut {NoStop}%
\bibitem [{\citenamefont {Guti{\'e}rrez}\ \emph
  {et~al.}(2019{\natexlab{b}})\citenamefont {Guti{\'e}rrez}, \citenamefont
  {Berrocal}, \citenamefont {Cornejo}, \citenamefont {Dom{\'i}nguez},
  \citenamefont {Del~Pozo}, \citenamefont {Arrazola}, \citenamefont
  {Ba\~nuelos}, \citenamefont {Escobedo}, \citenamefont {Lamata}, \citenamefont
  {Rica}, \citenamefont {Schmidt}, \citenamefont {Block}, \citenamefont
  {Solano},\ and\ \citenamefont {Rodr{\'i}guez}}]{Guti2019}%
  \BibitemOpen
  \bibfield  {author} {\bibinfo {author} {\bibfnamefont {M.~J.}\ \bibnamefont
  {Guti{\'e}rrez}}, \bibinfo {author} {\bibfnamefont {J.}~\bibnamefont
  {Berrocal}}, \bibinfo {author} {\bibfnamefont {J.~M.}\ \bibnamefont
  {Cornejo}}, \bibinfo {author} {\bibfnamefont {F.}~\bibnamefont
  {Dom{\'i}nguez}}, \bibinfo {author} {\bibfnamefont {J.~J.}\ \bibnamefont
  {Del~Pozo}}, \bibinfo {author} {\bibfnamefont {I.}~\bibnamefont {Arrazola}},
  \bibinfo {author} {\bibfnamefont {J.}~\bibnamefont {Ba\~nuelos}}, \bibinfo
  {author} {\bibfnamefont {P.}~\bibnamefont {Escobedo}}, \bibinfo {author}
  {\bibfnamefont {L.}~\bibnamefont {Lamata}}, \bibinfo {author} {\bibfnamefont
  {R.~A.}\ \bibnamefont {Rica}}, \bibinfo {author} {\bibfnamefont
  {S.}~\bibnamefont {Schmidt}}, \bibinfo {author} {\bibfnamefont
  {M.}~\bibnamefont {Block}}, \bibinfo {author} {\bibfnamefont
  {E.}~\bibnamefont {Solano}},\ and\ \bibinfo {author} {\bibfnamefont
  {D.}~\bibnamefont {Rodr{\'i}guez}},\ }\bibfield  {title} {\bibinfo {title}
  {The {TRAPSENSOR} facility: an open-ring 7 tesla {P}enning trap for
  laser-based precision experiments},\ }\href
  {https://doi.org/10.1088/1367-2630/aafa45} {\bibfield  {journal} {\bibinfo
  {journal} {New J. Phys.}\ }\textbf {\bibinfo {volume} {21}},\ \bibinfo
  {pages} {023023} (\bibinfo {year} {2019}{\natexlab{b}})}\BibitemShut
  {NoStop}%
\bibitem [{\citenamefont {Berrocal}\ \emph {et~al.}(2022)\citenamefont
  {Berrocal}, \citenamefont {Altozano}, \citenamefont {Dom\'inguez},
  \citenamefont {Guti\'errez}, \citenamefont {Cerrillo}, \citenamefont
  {Fern\'andez}, \citenamefont {Block}, \citenamefont {Ospelkaus},\ and\
  \citenamefont {Rodr\'iguez}}]{Berr2022}%
  \BibitemOpen
  \bibfield  {author} {\bibinfo {author} {\bibfnamefont {J.}~\bibnamefont
  {Berrocal}}, \bibinfo {author} {\bibfnamefont {E.}~\bibnamefont {Altozano}},
  \bibinfo {author} {\bibfnamefont {F.}~\bibnamefont {Dom\'inguez}}, \bibinfo
  {author} {\bibfnamefont {M.~J.}\ \bibnamefont {Guti\'errez}}, \bibinfo
  {author} {\bibfnamefont {J.}~\bibnamefont {Cerrillo}}, \bibinfo {author}
  {\bibfnamefont {F.~J.}\ \bibnamefont {Fern\'andez}}, \bibinfo {author}
  {\bibfnamefont {M.}~\bibnamefont {Block}}, \bibinfo {author} {\bibfnamefont
  {C.}~\bibnamefont {Ospelkaus}},\ and\ \bibinfo {author} {\bibfnamefont
  {D.}~\bibnamefont {Rodr\'iguez}},\ }\bibfield  {title} {\bibinfo {title}
  {Formation of two-ion crystals by injection from a {P}aul-trap source into a
  high-magnetic-field {P}enning trap},\ }\href
  {https://doi.org/10.1103/PhysRevA.105.052603} {\bibfield  {journal} {\bibinfo
   {journal} {Phys. Rev. A}\ }\textbf {\bibinfo {volume} {105}},\ \bibinfo
  {pages} {052603} (\bibinfo {year} {2022})}\BibitemShut {NoStop}%
\bibitem [{\citenamefont {Crick}\ \emph {et~al.}(2010)\citenamefont {Crick},
  \citenamefont {Donnellan}, \citenamefont {Segal},\ and\ \citenamefont
  {Thompson}}]{Cric2010}%
  \BibitemOpen
  \bibfield  {author} {\bibinfo {author} {\bibfnamefont {D.~R.}\ \bibnamefont
  {Crick}}, \bibinfo {author} {\bibfnamefont {S.}~\bibnamefont {Donnellan}},
  \bibinfo {author} {\bibfnamefont {D.~M.}\ \bibnamefont {Segal}},\ and\
  \bibinfo {author} {\bibfnamefont {R.~C.}\ \bibnamefont {Thompson}},\
  }\bibfield  {title} {\bibinfo {title} {Magnetically induced electron shelving
  in a trapped {C}a${}^{+}$ ion},\ }\href
  {https://doi.org/10.1103/PhysRevA.81.052503} {\bibfield  {journal} {\bibinfo
  {journal} {Phys. Rev. A}\ }\textbf {\bibinfo {volume} {81}},\ \bibinfo
  {pages} {052503} (\bibinfo {year} {2010})}\BibitemShut {NoStop}%
\bibitem [{\citenamefont {Powell}\ \emph {et~al.}(2002)\citenamefont {Powell},
  \citenamefont {Segal},\ and\ \citenamefont {Thompson}}]{Powe2002}%
  \BibitemOpen
  \bibfield  {author} {\bibinfo {author} {\bibfnamefont {H.~F.}\ \bibnamefont
  {Powell}}, \bibinfo {author} {\bibfnamefont {D.~M.}\ \bibnamefont {Segal}},\
  and\ \bibinfo {author} {\bibfnamefont {R.~C.}\ \bibnamefont {Thompson}},\
  }\bibfield  {title} {\bibinfo {title} {Axialization of laser cooled magnesium
  ions in a \uppercase{P}enning trap},\ }\href
  {https://doi.org/10.1103/PhysRevLett.89.093003} {\bibfield  {journal}
  {\bibinfo  {journal} {Phys. Rev. Lett.}\ }\textbf {\bibinfo {volume} {89}},\
  \bibinfo {pages} {093003} (\bibinfo {year} {2002})}\BibitemShut {NoStop}%
\bibitem [{\citenamefont {Wang}\ \emph {et~al.}(2020)\citenamefont {Wang},
  \citenamefont {Huang}, \citenamefont {Kondev}, \citenamefont {Audi},\ and\
  \citenamefont {Naimi}}]{Wang2020}%
  \BibitemOpen
  \bibfield  {author} {\bibinfo {author} {\bibfnamefont {M.}~\bibnamefont
  {Wang}}, \bibinfo {author} {\bibfnamefont {W.}~\bibnamefont {Huang}},
  \bibinfo {author} {\bibfnamefont {F.}~\bibnamefont {Kondev}}, \bibinfo
  {author} {\bibfnamefont {G.}~\bibnamefont {Audi}},\ and\ \bibinfo {author}
  {\bibfnamefont {S.}~\bibnamefont {Naimi}},\ }\bibfield  {title} {\bibinfo
  {title} {The \uppercase{AME} 2020 atomic mass evaluation (ii).
  \uppercase{T}ables, graphs and references},\ }\href
  {https://doi.org/10.1088/1674-1137/abddaf} {\bibfield  {journal} {\bibinfo
  {journal} {Chin. Phys. C}\ }\textbf {\bibinfo {volume} {45}},\ \bibinfo
  {pages} {030003} (\bibinfo {year} {2020})}\BibitemShut {NoStop}%
\bibitem [{\citenamefont {Bustabad}\ \emph {et~al.}(2013)\citenamefont
  {Bustabad}, \citenamefont {Bollen}, \citenamefont {Brodeur}, \citenamefont
  {Lincoln}, \citenamefont {Novario}, \citenamefont {Redshaw}, \citenamefont
  {Ringle}, \citenamefont {Schwarz},\ and\ \citenamefont
  {Valverde}}]{Bust2013}%
  \BibitemOpen
  \bibfield  {author} {\bibinfo {author} {\bibfnamefont {S.}~\bibnamefont
  {Bustabad}}, \bibinfo {author} {\bibfnamefont {G.}~\bibnamefont {Bollen}},
  \bibinfo {author} {\bibfnamefont {M.}~\bibnamefont {Brodeur}}, \bibinfo
  {author} {\bibfnamefont {D.}~\bibnamefont {Lincoln}}, \bibinfo {author}
  {\bibfnamefont {S.}~\bibnamefont {Novario}}, \bibinfo {author} {\bibfnamefont
  {M.}~\bibnamefont {Redshaw}}, \bibinfo {author} {\bibfnamefont
  {R.}~\bibnamefont {Ringle}}, \bibinfo {author} {\bibfnamefont
  {S.}~\bibnamefont {Schwarz}},\ and\ \bibinfo {author} {\bibfnamefont
  {A.}~\bibnamefont {Valverde}},\ }\bibfield  {title} {\bibinfo {title} {First
  direct determination of the $^{48}$\uppercase{C}a double-$\beta$ decay
  $\uppercase{Q}$ value},\ }\href {https://doi.org/10.1103/PhysRevC.88.022501}
  {\bibfield  {journal} {\bibinfo  {journal} {Phys. Rev. C}\ }\textbf {\bibinfo
  {volume} {88}},\ \bibinfo {pages} {022501(R)} (\bibinfo {year}
  {2013})}\BibitemShut {NoStop}%
\bibitem [{\citenamefont {Kwiatkowski}\ \emph {et~al.}(2014)\citenamefont
  {Kwiatkowski}, \citenamefont {Brunner}, \citenamefont {Holt}, \citenamefont
  {Chaudhuri}, \citenamefont {Chowdhury}, \citenamefont {Eibach}, \citenamefont
  {Engel}, \citenamefont {Gallant}, \citenamefont {Grossheim}, \citenamefont
  {Horoi}, \citenamefont {Lennarz}, \citenamefont {Macdonald}, \citenamefont
  {Pearson}, \citenamefont {Schultz}, \citenamefont {Simon}, \citenamefont
  {Senkov}, \citenamefont {Simon}, \citenamefont {Zuber},\ and\ \citenamefont
  {Dilling}}]{Kwia2014}%
  \BibitemOpen
  \bibfield  {author} {\bibinfo {author} {\bibfnamefont {A.}~\bibnamefont
  {Kwiatkowski}}, \bibinfo {author} {\bibfnamefont {T.}~\bibnamefont
  {Brunner}}, \bibinfo {author} {\bibfnamefont {J.~D.}\ \bibnamefont {Holt}},
  \bibinfo {author} {\bibfnamefont {A.}~\bibnamefont {Chaudhuri}}, \bibinfo
  {author} {\bibfnamefont {U.}~\bibnamefont {Chowdhury}}, \bibinfo {author}
  {\bibfnamefont {M.}~\bibnamefont {Eibach}}, \bibinfo {author} {\bibfnamefont
  {J.}~\bibnamefont {Engel}}, \bibinfo {author} {\bibfnamefont {A.~T.}\
  \bibnamefont {Gallant}}, \bibinfo {author} {\bibfnamefont {A.}~\bibnamefont
  {Grossheim}}, \bibinfo {author} {\bibfnamefont {M.}~\bibnamefont {Horoi}},
  \bibinfo {author} {\bibfnamefont {A.}~\bibnamefont {Lennarz}}, \bibinfo
  {author} {\bibfnamefont {T.~D.}\ \bibnamefont {Macdonald}}, \bibinfo {author}
  {\bibfnamefont {M.~R.}\ \bibnamefont {Pearson}}, \bibinfo {author}
  {\bibfnamefont {B.~E.}\ \bibnamefont {Schultz}}, \bibinfo {author}
  {\bibfnamefont {M.~C.}\ \bibnamefont {Simon}}, \bibinfo {author}
  {\bibfnamefont {R.~A.}\ \bibnamefont {Senkov}}, \bibinfo {author}
  {\bibfnamefont {V.~V.}\ \bibnamefont {Simon}}, \bibinfo {author}
  {\bibfnamefont {K.}~\bibnamefont {Zuber}},\ and\ \bibinfo {author}
  {\bibfnamefont {J.}~\bibnamefont {Dilling}},\ }\bibfield  {title} {\bibinfo
  {title} {New determination of double-$\beta $-decay properties in
  $^{48}$\uppercase{C}a: High-precision $\uppercase{Q}_{\beta \beta}$-value
  measurement and improved nuclear matrix element calculations},\ }\href
  {https://doi.org/10.1103/PhysRevC.89.045502} {\bibfield  {journal} {\bibinfo
  {journal} {Phys. Rev. C}\ }\textbf {\bibinfo {volume} {89}},\ \bibinfo
  {pages} {045502} (\bibinfo {year} {2014})}\BibitemShut {NoStop}%
\bibitem [{\citenamefont {K\"ohler}\ \emph {et~al.}(2016)\citenamefont
  {K\"ohler}, \citenamefont {Blaum}, \citenamefont {Block}, \citenamefont
  {Chenmarev}, \citenamefont {Eliseev}, \citenamefont {Glazov}, \citenamefont
  {Goncharov}, \citenamefont {Hou}, \citenamefont {Kracke}, \citenamefont
  {Nesterenko}, \citenamefont {Novikov}, \citenamefont {Quint}, \citenamefont
  {Minaya-Ramirez}, \citenamefont {Shabaev}, \citenamefont {Sturm},
  \citenamefont {Volotka},\ and\ \citenamefont {Werth}}]{Kohl2016}%
  \BibitemOpen
  \bibfield  {author} {\bibinfo {author} {\bibfnamefont {F.}~\bibnamefont
  {K\"ohler}}, \bibinfo {author} {\bibfnamefont {K.}~\bibnamefont {Blaum}},
  \bibinfo {author} {\bibfnamefont {M.}~\bibnamefont {Block}}, \bibinfo
  {author} {\bibfnamefont {S.}~\bibnamefont {Chenmarev}}, \bibinfo {author}
  {\bibfnamefont {S.}~\bibnamefont {Eliseev}}, \bibinfo {author} {\bibfnamefont
  {D.~A.}\ \bibnamefont {Glazov}}, \bibinfo {author} {\bibfnamefont
  {M.}~\bibnamefont {Goncharov}}, \bibinfo {author} {\bibfnamefont
  {J.}~\bibnamefont {Hou}}, \bibinfo {author} {\bibfnamefont {A.}~\bibnamefont
  {Kracke}}, \bibinfo {author} {\bibfnamefont {D.}~\bibnamefont {Nesterenko}},
  \bibinfo {author} {\bibfnamefont {Y.~N.}\ \bibnamefont {Novikov}}, \bibinfo
  {author} {\bibfnamefont {W.}~\bibnamefont {Quint}}, \bibinfo {author}
  {\bibfnamefont {E.}~\bibnamefont {Minaya-Ramirez}}, \bibinfo {author}
  {\bibfnamefont {V.}~\bibnamefont {Shabaev}}, \bibinfo {author} {\bibfnamefont
  {S.}~\bibnamefont {Sturm}}, \bibinfo {author} {\bibfnamefont
  {A.}~\bibnamefont {Volotka}},\ and\ \bibinfo {author} {\bibfnamefont
  {G.}~\bibnamefont {Werth}},\ }\bibfield  {title} {\bibinfo {title} {Isotope
  dependence of the \uppercase{Z}eeman effect in lithium-like calcium},\ }\href
  {https://doi.org/10.1038/ncomms10246} {\bibfield  {journal} {\bibinfo
  {journal} {Nature Comm.}\ }\textbf {\bibinfo {volume} {7}},\ \bibinfo {pages}
  {10246} (\bibinfo {year} {2016})}\BibitemShut {NoStop}%
\bibitem [{\citenamefont {Ketter}\ \emph {et~al.}(2014)\citenamefont {Ketter},
  \citenamefont {Eronen}, \citenamefont {H\"ocker}, \citenamefont {Streubel},\
  and\ \citenamefont {Blaum}}]{Kett2014}%
  \BibitemOpen
  \bibfield  {author} {\bibinfo {author} {\bibfnamefont {J.}~\bibnamefont
  {Ketter}}, \bibinfo {author} {\bibfnamefont {T.}~\bibnamefont {Eronen}},
  \bibinfo {author} {\bibfnamefont {M.}~\bibnamefont {H\"ocker}}, \bibinfo
  {author} {\bibfnamefont {S.}~\bibnamefont {Streubel}},\ and\ \bibinfo
  {author} {\bibfnamefont {K.}~\bibnamefont {Blaum}},\ }\bibfield  {title}
  {\bibinfo {title} {First-order perturbative calculation of the
  frequency-sifts caused by static cylindrically-symmetric electric and
  magnetic imperfections of a \uppercase{P}enning trap},\ }\href
  {https://doi.org/10.1016/j.ijms.2013.10.005} {\bibfield  {journal} {\bibinfo
  {journal} {Int. J. Mass Spectrom.}\ }\textbf {\bibinfo {volume} {358}},\
  \bibinfo {pages} {1} (\bibinfo {year} {2014})}\BibitemShut {NoStop}%
\bibitem [{\citenamefont {Redshaw}(2007)}]{Reds2007}%
  \BibitemOpen
  \bibfield  {author} {\bibinfo {author} {\bibfnamefont {M.}~\bibnamefont
  {Redshaw}},\ }\emph {\bibinfo {title} {Precise Measurements of the Atomic
  Masses of $^{28}$Si, $^{31}$P, $^{32}$S, $^{84,86}$Kr, $^{129,132,136}$Xe,
  and the Dipole Moment of Ph$^+$ Using Single-Ion and Two-Ion Penning Trap
  Techniques}},\ \href@noop {} {Ph.D. thesis},\ \bibinfo  {school} {Florida
  State University} (\bibinfo {year} {2007})\BibitemShut {NoStop}%
\bibitem [{\citenamefont {Sch\"ussler}(2019)}]{Schu2019}%
  \BibitemOpen
  \bibfield  {author} {\bibinfo {author} {\bibfnamefont {R.}~\bibnamefont
  {Sch\"ussler}},\ }\emph {\bibinfo {title} {First High-Precision Mass
  Measurements at PENTATRAP on highly charged Xe and Re ions}},\ \href@noop {}
  {Ph.D. thesis},\ \bibinfo  {school} {University of Heidelberg} (\bibinfo
  {year} {2019})\BibitemShut {NoStop}%
\end{thebibliography}

\providecommand{\noopsort}[1]{}\providecommand{\singleletter}[1]{#1}%

\end{document}